\documentstyle[amssymb,aps]{revtex}

\begin{document}
\author{A.L. Cornelius$^{1}$ and J.S. Gardner$^{2}$}
\address{$^{1}$Department of Physics, University of Nevada Las Vegas,\\
Las Vegas, Nevada, 89154-4002}
\address{$^{2}$National Research Council, NPMR, Chalk River Laboratories,\\
Chalk River, Ontario, Canada K0J 1J0}
\title{Short-range Magnetic interactions in the Spin-Ice compound \\
Ho$_{2}$Ti$_{2}$O$_{7}$}
\date{\today }
\maketitle

\begin{abstract}
Magnetization and susceptibility studies on single crystals of the
pyrochlore Ho$_{2}$Ti$_{2}$O$_{7}$ are reported for the first time.
Magnetization isotherms are shown to be qualitatively similar to that
predicted by the nearest neighbor spin-ice model. Below the lock-in
temperature, $T^{\ast }\simeq 1.97$ K, magnetization is consistent with the
locking of spins along [111] directions in a specific two-spins-in,
two-spins-out arrangement. Below $T^{\ast }$ the magnetization for $B||[111]$
displays a two step behavior signalling the breaking of the ice rules.
\end{abstract}


\twocolumn[\hsize\textwidth\columnwidth\hsize\csname@twocolumnfalse\endcsname

\twocolumn\vskip.25pc]\narrowtext

Materials containing antiferromagnetically coupled magnetic moments which
reside on geometrical units, such as triangles, that inhibit the formation
of a collinear ordered state, often display phenomena known broadly as
geometrical frustration \cite{review}. A pyrochlore lattice with
corner-sharing tetrahedra of magnetic ions running through it, is an ideal
system for studying geometrical frustration. The discovery by Harris {\it et
al}.~\cite{Harris97} that a magnetic pyrochlore with {\it ferromagnetic}
interactions and strong single ion anisotropy could be just as frustrated as
the analogous antiferromagnets is counterintuitive \cite%
{Moessner98,Bramwell98}. The prototypical example of these ferromagnetic
compounds is Ho$_{2}$Ti$_{2}$O$_{7}$~\cite{Harris97,Harris98,Bramwell01},
however this has been questioned recently~\cite{Siddharthan99}. Such
materials are termed spin-ices because their physics maps directly onto the
historic problem of proton disorder in hexagonal ice.

The magnetic ground state of the ferromagnetic pyrochlore with Ising
anisotropy along the $[111]$ axis consists of two spins pointing in and two
spins pointing out of the individual tetrahedra \cite{Harris97}. This is
identical to the lattice formed by the mid-points of the oxygen-oxygen bonds
in ice~\cite{Anderson56}. This resulted in the ``spin-ice'' label, and is
one of two conditions required for this class of materials, the other being
a zero-point entropy close to $(1/2)\ln (3/2)$ which has been measured in Dy$%
_{2}$Ti$_{2}$O$_{7}$, and is identical to the proton ordering in real ice.
The discovery of spin-ice prompted considerable research by other groups~%
\cite{Siddharthan99,Ramirez99,Hertog00}, however until recently neither Dy$%
_{2}$Ti$_{2}$O$_{7}$ nor Ho$_{2}$Ti$_{2}$O$_{7}$ have been shown to meet
these two criteria. Due to the neutron absorbing nature of dysprosium,
neutron scattering measurements on Dy$_{2}$Ti$_{2}$O$_{7}$ to investigate
the nature of the magnetic ordering have yet to be performed. Recently,
Bramwell {\it et al.} reported the first low temperature $C_{m}$ results on
Ho$_{2}$Ti$_{2}$O$_{7}$~\cite{Bramwell01}. In these data a maximum is seen
in the magnetic contribution, $C_{m}$, to the specific heat at 1.97 K which
we will refer to as the blocking temperature. Their data, heat capacity and
neutron diffraction, clearly shows that Ho$_{2}$Ti$_{2}$O$_{7}$ requires a
spin-ice model that includes longer-range, dipolar interactions when
discussing its properties down to 300 mK.

To investigate the bulk magnetic properties of Ho$_{2}$Ti$_{2}$O$_{7}$, we
have performed both a.c. and d.c. magnetization measurements on single
crystals. These are the first single crystal measurements, with fields as
large as 9 T applied along the three principle cubic crystallographic axes $%
[100],$ $[110]$ and $[111]$, of this geometrically frustrated materials.
These data are in excellent qualitative agreement with calculations \cite%
{Harris98}; in particular a step is seen below the lock-in temperature, $%
T^{\ast }$ when the magnetic field is applied along the $[111]$ axis. Unlike
the predictions, there is little difference in the low temperature
saturation magnetization for the different crystal orientations.

The single crystals were synthesized using the floating-zone method
described elsewhere \cite{Gardner98}. Polycrystalline samples of Ho$_{2}$Ti$%
_{2}$O$_{7}$ were prepared by firing, in air, stoichiometric amounts of high
purity ($>99.99\%$) Ho$_{2}$O$_{3}$ and TiO$_{2}$ for several days with
intermittent grindings to ensure a complete reaction. The polycrystalline
material is then used to grow large ($>1$ cm$^{3}$) single crystals in an
infrared image furnace with the aid of a small seed crystal. A small piece
of the resulting large single crystal was ground up for characterization by
X-ray powder diffraction (XRD). The impurity content is less than 3\% (limit
of our XRD) and the room temperature cubic lattice parameter is $10.102(2)$
\AA.

Magnetization measurements between 1.8 and 300~K were performed in a fully
calibrated Quantum Design PPMS\ system inside a 9 T\ superconducting magnet.
The zero field susceptibility measurements used a 10 Oe AC\ field at a
frequency of 1000 Hz. The high field magnetization measurements were made
with a DC\ extraction technique.

Previous magnetic measurements on Ho$_{2}$Ti$_{2}$O$_{7}$ have been
interpreted in different ways \cite{Siddharthan99,Bramwell00,Jana00}. We
will discuss our results within the framework of a ground-state spin
one-half doublet $S=1/2$ with an effective $g$ value of $g_{\parallel }=g$
and $g_{\perp }=0$ where parallel is the $[111]$\ direction corresponding to
the direction of the applied field. We will use the standard form \cite%
{Ashcroft76} 
\begin{equation}
\chi (T)=\frac{\left( g\mu _{B}\right) ^{2}}{k_{B}T}S(S+1)\left[ 1+\frac{%
\theta }{T}\right] ,\theta =S(S+1)J/3,
\end{equation}%
where $J$ is the effective interaction between the Ho spins. As noted by
others \cite{Gingras00}, $\theta $ is going to be the sum of the exchange,
dipole, and crystal field contributions. We do not attempt to extract these
individual contributions to $\theta $. The results of the AC zero field
susceptibility measurements are shown in Fig. \ref{ho susc}. For $T>10$ K,
the data are consistent with that expected for $g=20$ and weak ferromagnetic
interactions as previous report~\cite{Bramwell00}. In the range 3-10~K, the
data can be described well using $\theta =-1.8$~K which corresponds to an
antiferromagnetic interaction, in agreement with the results of Siddharthan 
{\it et al}. \cite{Siddharthan99}. Jana {\it et al.}~\cite{Jana00} and other
have already noted that the Curie-Weiss temperature is very dependent on the
range of temperatures studied, this is undoubtedly due to crystal field
contributions to the effective interaction. Rosenkranz ${\it {et~al.}}$\cite%
{Rosenkranz00} have shown that the lowest crystal field excitation is at $%
\sim $ 200 K. If we fit our high temperature, $T~>~200$ K, DC\ magnetization
data, (Fig. \ref{static}), with $B||[111]$=1 kOe, we find $\theta \approx
+1.6$ K, in agreement with others \cite{Bramwell00}. From these results, we
conclude that the Ho moments can indeed be considered as $S=1/2$ spins, with 
$g=20$ and weak, but dominant ferromagnetic interactions in the paramagnetic
state. It should be noted there that this net ferromagnetic interaction
between Ho spins is a combination of antiferromagnetic exchange interactions
and ferromagnetic dipolar coupling.

Although not shown here, it can be shown that $C_{m}$ is also well described
by the nearest neighbor spin-ice model at temperatures above 1K. Using the
data from Bramwell ${\it {etal.}}$~\cite{Bramwell01} we have calculated the
magnetic entropy $\Delta S_{m}$ by integrating $C_{m}/T$ and the results are
shown in the inset of Fig. \ref{static}. The error bars represent the
estimated uncertainty in the entropy, mainly due to the fit of hyperfine
component. The solid line is the entropy expected from Harris {\it et al}.~%
\cite{Harris98}. The behavior is remarkably similar to that observed in Dy$%
_{2}$Ti$_{2}$O$_{7}$ \cite{Ramirez99} where the application of a 1 T
magnetic field nearly recovers the missing $(1/2)R\ln (3/2)$ of entropy
expected for a spin 1/2 system ($R\ln 2$). This is in sharp contrast to a
previous report \cite{Siddharthan99}, and is consistent with Ho$_{2}$Ti$_{2}$%
O$_{7}$ being a spin-ice compound.

One naively expects magnetic ordering for $T<|\theta |$. However, the
geometric frustration leads to the absence of magnetic order at least down
to 0.05 K \cite{Harris97}. Though long-range magnetic order does not appear,
the magnetic interactions can lead to short-range local ordering.
Measurements of the DC\ magnetization as a function of magnetic field are
shown in Fig. \ref{ho mag}. Before each of these measurements, the
temperature was raised to 10 K in zero applied field before returning to the
desired temperature and applying a magnetic field. For $T>2$ K, the curves
for $B||[111]$ and $B||[100]$ are identical, while for $T\lesssim 2$ K
drastically different behavior is observed. The data for the $B||[110]$
orientation is rather interesting. While the other two orientations show
identical behavior for $T>T^{\ast }$, suggestive of isotropic magnetization,
the $B||[110]$ data is rather different. Also, for $T<T^{\ast }$ the $%
B||[110]$ data does not differ appreciably from the 4 K\ data. In other
words, the locking of the spins that we believe to occur for $T<T^{\ast }$
seems to have little effect on the behavior when $B||[110]$. It should be
noted that when $B||[110]$, two of the four spins are perpendicular to the
field and cannot order. In the other directions all spins have a component
of their moment in the field direction. In general the data can be explained
by the spins having strong Ising anisotropy along the $[111]$ axis. Below $%
T^{\ast }\sim 1.9$ K$\sim |\theta |$ the spins lock into a specific set of
[111] direction so that a tetrahedron has the two-spins-in, two-spins-out
configuration. From Monte Carlo calculations of the nearest neighbor
spin-ice lattice \cite{Harris98}, it was shown that the degeneracy breaking
occurs via a two step process for $B||[111]$. First the spin pointing along
the $[111]$ direction is aligned with the field leading to an average moment
in the field direction of (1+1/3+1/3-1/3)/4=1/3 followed by a breaking of
the two-spins-in two-spins-out ice rules with an average moment of
(1+1/3+1/3+1/3)/4=1/2 at higher fields. This predicted behavior yields a
plateau at 3.33 $\mu _{\text{B}}$/Ho atom followed by saturation at 5.00 $%
\mu _{\text{B}}$/Ho atom for the case $g=20$. Our data for $B||[111]$
clearly shows a plateau at $\sim 3.3$ $\mu _{\text{B}}$/Ho appearing as
temperature is lowered, which agrees well with the predicted behavior.
However, the saturation magnetization is approximately 20\% higher than
calculated. For $B||[100]$ and $B||[110]$, the data at $T=1.8$~K
qualitatively agree well with the predicted behavior \cite{Harris98}. In
particular, for $T<T^{\ast }$, the field required to reach saturation is
significantly less for $B||[100]$ as for $B||[110]$. It is seen that, in
stark contrast to the predicted behavior of a saturation moment of $5.78$ $%
\mu _{\text{B}}$/Ho for $B||[100]$ and $4.08$ $\mu _{\text{B}}$/Ho for $%
B||[110]$, the measured saturation moment of $5.9$ $\mu _{\text{B}}$/Ho is
close to the value expected for $B||[100]$ and is nearly independent of
direction, though it requires a higher field than reached in the current
experiment to say for certainty that saturation has been achieved for $%
B||[110]$. We have no definitive explanation for the lack of anisotropy in
the measured saturation moment, perhaps the influence of the dipolar
interaction, which is not taken into account by Harris {\it et al}. \cite%
{Harris98}, plays a role. We should also point out that the calculations of
Harris {\it et al}. \cite{Harris98} are at $T/J=0.1\sim 0.3K$, which is much
lower than our lowest temperature, but the qualitative agreement is good
since both temperatures are below $T^{\ast }$. This is the same conclusion
of a different simulation that shows a single step like feature followed by
saturation for $B||[111]$ and an identical saturation moment (though a much
different $M$ versus $B$ curve) for a different field direction \cite%
{Siddharthan00}.

We have performed the first measurements of the magnetization on Ho$_{2}$Ti$%
_{2}$O$_{7}$ single crystals along various crystalline directions. The
magnetization results can be explained by $S=1/2$ spins with $g=20$ which
lock in along the [111] direction at $T^{\ast }\approx 1.97$ K, and the
results of applying the field along the different crystalline axes are
explained qualitatively by a Monte Carlo calculation of the spin-ice model
with nearest neighbor interactions ${\it {only}}$~\cite{Harris98}. In
particular, the observation of a plateau in the magnetization as a function
of applied magnetic field for $B||[111]$ at $T<T^{\ast }$ as predicted \cite%
{Harris98}, has been observed for the first time. Contrary to predicted
behavior \cite{Harris98}, it was found that the saturation magnetic moment
of $\sim 5.9$ $\mu _{\text{B}}$/Ho is nearly independent of the orientation
of the crystalline axes relative to the applied field. The magnetization and
heat capacity measurements are both consistent with an interaction strength
of $J\approx 7.2-7.4$ K between the Ho magnetic moments.

In conclusion, above $T^{\ast }$ the spins have a preferred orientation
along the [111] direction due to single ion anisotropy revealed by the
reduced saturated moment in the system. From magnetization measurements
below 2 K the moments on a single tetrahedron have a specific two-spins-in,
two-spins-out configuration and the moments cannot overcome an energy
barrier to another configuration. This result is consistent with neutron
scattering work which finds the Ho spins in a two-spins-in, two-spins-out
configuration and recent specific heat results~\cite{Bramwell01}.

One of us (J.G.) would like to thank S. Bramwell, J. Greedan and A. Ramirez
for fruitful discussions. This work was supported by DOE Materials Science
Contract No. DE-FG02-00ER45835 and Cooperative Agreement DE-FC08-98NV13410
(A.C.).


\begin{figure}[tbp]
\caption{AC\ susceptibility times temperature versus inverse temperature on
a Ho$_{2}$Ti$_{2}$O$_{7}$ single crystal for magnetic field applied along
the $[111]$ axis. The horizontal line is the expected value for $g=20$. The
other line is a fit described in the text.}
\label{ho susc}
\end{figure}

\begin{figure}[tbp]
\caption{The inverse static susceptibility in an applied field of 1 kOe as a
function of temperature. The line is a Curie-Weiss fit as described in text.
The inset shows the magnetic entropy $\Delta S_{m}$ as a function of
temperature $T$ for Ho$_{2}$Ti$_{2}$O$_{7}$ at 0 T and 1 T. The solid line
corresponds to the expected values of the nearest neighbor spin-ice model as
discussed in the text. The dashed lines represent the recovery of $(1/2)R\ln
(3/2)$ of the expected $R\ln 2$ entropy as discussed in text.}
\label{static}
\end{figure}

\begin{figure}[tbp]
\caption{Magnetization as a function of applied magnetic field along the
listed crystalline directions for Ho$_{2}$Ti$_{2}$O$_{7}$ at various
temperatures. A plateau is clearly observed in the data for $B||[111]$ as
the temperature is lowered below $T^{\ast }\simeq 2$ K.}
\label{ho mag}
\end{figure}

\end{document}